\documentstyle[times,pramana,epsf,floats]{ias}

\begin{document}
\mark{{Possible Neutrino Detector in India}{M. V. N. Murthy et al}}
\title{Discussion on a possible neutrino detector located in India }
\author{{\it Coordinators} : M. V. N.~MURTHY${}^{(1)}$ and U.
A.~YAJNIK${}^{(7)}$
\\
\\
{\it Participants and contributing authors}:
K.R.S. Balaji (1), G. Bhattacharyya (2), Amol Dighe (3), Shashikant Dugad
(4), N.D. Hari Dass (1), P.K. Kabir (5), Kamales Kar (2), D. Indumathi
(1), John G. Learned (6), Debasish Majumdar (2), N.K. Mondal (4),
M.V.N. Murthy (1), S.N. Nayak (7), Sandip Pakvasa (6), Amitava
Raychaudhuri (8), R.S. Raghavan (9), G. Rajasekaran (1), R. Ramachandran
(1), Alak K. Ray (4), Asim K. Ray (10), Saurabh Rindani (11),
H.S. Sharatchandra (1), Rahul Sinha (1), Nita Sinha (1), S. Umasankar
(7), Urjit A. Yajnik (7)}
\address
{ (1) IMSc, Chennai
(2) SINP, Calcutta
(3) CERN, Geneva
(4) TIFR, Mumbai
(5) U. Virginia, Charlottesville
(6) U. Hawaii, Honolulu
(7) IIT Bombay, Mumbai 
(8) U. Calcutta, Calcutta
(9) Bell Labs, Lucent Tech, Murray Hill
(10) Viswa Bharati, Santiniketan
(11) PRL, Ahmedabad}

\abstract{ We have identified some important and worthwhile physics
opportunitites with a possible neutrino detector located in India.
Particular emphasis is placed on the geographical advantage with a
stress on the complimentary aspects with respect to other neutrino
detectors already in operation.}
 \maketitle

\section{Motivation}

\begin{itemize}

\item The possibility of a Neutrino Observatory located in India was
discussed as early as 1989 during several meetings held that year.
Since then this question has come up in many private discussions off
and on. The issue was raised again in the  first meeting of the
Neutrino physics and Cosmology working group during WHEPP-6 and it was
decided then to collate concrete ideas for such a detector.

\item Historically, the Indian initiative in Cosmic Ray and Neutrino
physics experiments goes back nearly 45 years. In fact the first
atmospheric neutrino induced muon events were recorded at the KGF
underground laboratory nearly thirty five years back. While Cosmic Ray
experiments  are still continuing, Indian neutrino physics has taken a
beating due to the  closure of one of the best locations, Kolar Gold
Mines, for such experiments. It appears that a resurrection is
possible only if there are alternative  locations available.

\item The biggest advantage in such a detector is the geographical
location. Most of the neutrino detectors are scattered around the
world  at latitudes above 35$^o$. There is none close to the equator
as yet.  It is possible to push such a detector (henceforth called INO) 
down to almost 8$^o$ latitude if a
convenient location is available. Equatorial location permits solar
neutrino oscillation studies through the earth's core, possible
studies of the core density, and permits doing neutrino astronomy
searches covering the whole celestial sky.

\item The working group concentrated on generating physics ideas for
such an initiative without necessarily specifying what type of
detector is necessary for studying the physics. However, parallel
discussions also took place on some specific configurations guided by
the physics  discussions. The major motivation behind this exercise is
then to detail all the relevant physics ideas, and then to view them
in juxtaposition with the available experimental resources and
expertise. Having thus highlighted the possible goals of such an
Indian Neutrino Observatory, it may then become easier to home-in on
one or more suitable experiments, both in the long-term and in the
short-term, which can be carried on there.

\item An important aspect of INO should also be that not only should
it have new physics ideas, it should be complementary to the other
detectors which are already running, or will be on-line soon. Such a
world class facility is naturally expected to attract worldwide
attention as well as participation. It may also have continuous
operations such as background measurements of various kinds which,
while not being fashionable, will also help in realising front line
and secondary physics objectives.
 
\end{itemize}

\section{Physics Goals}

This is best studied by considering different aspects of neutrino physics 
separately. All these ideas may not be incorporated in a single 
detector. Rather this is more like a wish list. 

\subsection{Solar neutrino physics}

The current and planned solar neutrino detectors are all at latitudes
above 35 degrees. These detectors have very small exposure times to
neutrinos traveling through the core of the earth, about ten days in
a year.  The MSW phenomenon can play an important role in the
regeneration of neutrinos in earth's core. To study this, it is
important to build a solar neutrino detector as close to the equator
as possible. India offers a possibility geographically, as well as in
terms of scientific expertise, engineering expertise, manufacturing,
and tall mountains which provide good shielding.

Such a detector will allow one to check the parameter range for
time-of-night variation. Some effort is  needed to examine the
theoretical issues involved  more closely. For example, there is a
need to do a detailed calculation of rates, for a hypothetical detector
located close to the equator, taking into account the constraints on the
neutrino parameters  already in existence.  This effort is needed to
examine the basis of justification of such a large endeavour, but
preliminary indications at this workshop make it appear promising.

While many aspects of solar neutrinos will have been well investigated by
the time an Indian effort can be mounted, there are some further areas
which will remain uninvestigated.  An example would be the measurement of
the higher energy tail of the solar neutrinos, in the range of 13-17 MeV
where Hep neutrinos can be measured, and where SuperK, the presently
largest instrument planned, will not have accumulated good statistics.
Typical number of events at a 1 kTon water Cerenkov detector are given in
Appendix A. The effects of cuts on the minimum energy of detection (of the
recoil electron) are shown in Table 1 in the Appendix. 

\subsection{Neutrinos from Stellar Collapse}

Typically in a type II supernova explosion, neutrinos and antineutrinos of
all types are emitted. The electron type neutrinos are expected to have an
average energy of about 12 MeV with a Fermi distribution. The electron
antineutrinos have an average energy of 16 MeV, where as the other species
have a higher average of about 25 MeV. The energy spectrum is
approximately given by the Fermi distribution with highest energies in the
range of about 80 to 100 MeV. 

Typically in a water Cerenkov detector, the charged current excitation of
the oxygen nucleus contributes only a small percentage of events if one
assumes that the neutrino flux is unchanged at the detector as compared to
the flux at source. This is primarily due to the high threshold of about
15~MeV in CC excitation in the oxygen nucleus. In the presence of mixing,
after imposing the known constraints, the $\nu_e$ and $\nu_{\mu,\tau}$
spectra are interchanged. As a result, a part or whole of $\nu_{\mu,\tau}$
flux appears as the $\nu_e$ flux at the detector. Because the original
$\nu_{\mu,\tau}$ spectrum is hotter, it leads to increased number of CC
nuclear excitations.  Using the spectral information from the supernova
neutrino emission, the increase may even be as large as a factor of 20 in
the $\nu_e$ CC sector. See Table 2 and Appendix B for details of events in
all sectors. Here, the effects of cuts on the recoil electron energy are
also shown. Interestingly enough the neutrino- (and antineutrino-) nucleus
CC events are backward peaked with respect to the supernova direction
unlike the neutrino (antineutrino) electron scattering events which show a
marked preference to the forward direction.  While mixing does not deplete
the forward peaked events on electron targets (in fact it increases
slightly because of the increase in average energy), it results in a
remarkable enhancement of the backward peaked events.  This signal is
model independent since the isotropic events due to $\bar\nu_e p
\rightarrow e^+ n$ events provide a large overall normalisation flux. 

Given that the nearest galaxy Andromeda is about 700 Kpc away, any
experiment which aims to detect neutrinos from a supernova event every few
years should have a sensitivity to probe distances of the order of 2-3 Mpc
assuming that a core collapse takes place in a typical galaxy once in
fifty years. This immediately puts the active mass of the detector system
in the megaton range given the small cross section of neutrinos of
energies greater than 10 MeV.  Since only one supernova, namely SN1987A
has been detected through neutrinos, a data set of approximately one SN
per year accumulated over several decades will be of enormous importance
to stellar astrophysics. 

\subsection{Short and Long Baseline Experiments with Neutrino Factories}

Muon storage rings are being seriously considered as a first step to a
$\mu^+\mu^-$ collider. The neutrinos are emitted in pairs $\nu_{\mu},
\bar\nu_{e}$ or $\bar\nu_{\mu}, \nu_{e}$ from $\mu^-$ or $\mu^+$ decays. 
Neutrino beams resulting from muon decay will have a precisely known flux,
well defined energy spectrum, and unprecedented intensity.  Very long
baseline neutrino oscillation experiments become possible with such
neutrino sources. Furthermore, they can test whether the dominant
atmospheric neutrino oscillations are $\nu_{\mu} \rightarrow \nu_{\tau}$
and/or $\nu_{\mu} \rightarrow \nu_{s}$ (sterile), in addition to
determining the $\nu_{\mu} \rightarrow \nu_e$ content of atmospheric
neutrino oscillations, and measure the $\nu_e \rightarrow \nu_{\tau}$
appearance. Depending on the oscillation parameters, they may be able to
detect earth matter and CP violation effects and to determine the ordering
of some of the mass eigenstates. An important feature of a detector for
this application is muon charge identification in the detector (if
possible). 

Several possibilities for the baseline experiments are already under
consideration. Two of these possibilities are Fermilab to MINOS (soon to
operate with a traditional neutrino source) and Fermilab to Gran Sasso
which is under consideration for later application with a neutrino
factory. In the first case the base-length will be 732 kms.  The
anticipated number of charged current events at this distance is nearly
20,000 per year with the neutrino factory. There are also proposals to
have baseline experiments from CERN to Gran Sasso (OPERA and ICANOE). 

To disentangle matter effects one needs data from two experiments with
different base lengths. An INO will have a baseline of about 10,000 kms
from FermiLab and the data from this experiment will be extremely useful
in determining neutrino parameters. This is perhaps the longest base
length one can envisage in addition to providing reasonably dense matter
in-between, during the long passage through the earth. Such a set-up will
have a geometry where the neutrinos barely graze the core; hence
core-matter effects may not be large. The experiment from Fermilab to Gran
Sasso misses the core entirely.  The combination of different baselines to
INO, for example from CERN and/or KEK in addition to Fermilab, will
provide leverage for unentangling the oscillation solutions. 

Another attractive feature of this proposal is that the storage rings will
perhaps be built in about ten years time. This is also the approximate
time needed to build a neutrino detector including planning, building,
calibration, etc. Unlike other proposals, this will be a completely new,
and front line experiment without the risk of being outdated even before a
beginning is made.  By necessity, this proposal must become a part of an
international setup.  as it involves availing the beam from a remote
source,

\subsection{Global radioactivity in the earth}

The antineutrino $\bar\nu_e$ spectroscopy can be used to measure the
separate global abundances of $^{238}$U and $^{232}$Th. The total internal
heat in the earth is of the order of 40 Terra Watts (TW). Of this about 40
percent or 16 TW is supposed to be due to the radiogenic heat. Models of
the earth disperse this heat equally between the core and the mantle. 
Within the mantle the heat is generated mainly in the continental crust
and a much smaller amount in the oceanic crust. A practical way of
studying this distribution is to detect antineutrinos through the reaction
$\bar\nu_e p \rightarrow e^+n$ since all the antineutrinos essentially
reach the surface. The present sub-MeV threshold neutrino detectors can
effectively be used to detect the neutrinos. However, the complete mapping
of the origin of radiogenic heat, hence the detailed structure of earth
matter, depends on putting an array of such detectors at various locations
with contrasting geological features. The conceptual foundations of the
geophysical structure, dynamics and evolution of the earth depends on such
detailed information. 

\subsection{Nucleon Decay}

One of the most important quests in all of physics remains the search for
nucleon decay.  The observation (or lack thereof) provides tests of
elementary particle structure at the unification scale, not accessible to
accelerator experiments.  The present best limits on the mode of $p
\rightarrow e^+ \pi^o$ are at a lifetime of $3 \times 10^{34}$ years
(from SuperK). These may be pushed to $10^{35}$ years. Current
predictions from theory range upwards by several orders of magnitude, but
nucleon decay could well be found at less than $10^{35}$ years. Many SUSY
models unfortunately favour nucleon decay modes with $K$'s, which have
atmospheric neutrino background, and for which SK will probably not do
much better than $10^{34}$ years. 

A megaton detector contains $6 \times 10^{35}$ nucleons.  Given such a
detector and the forthcoming more precise determinations of atmospheric
neutrino fluxes and cross sections, it should be possible to seriously
confront models with such an instrument (we discuss one realization
below). 

\subsection{Others}

Most of the physics goals mentioned above need underground laboratories
in order to reduce the background. No matter what the main physics goal
is, it is always possible to incorporate changes/additions such that
one can also look at outstanding problems. One such problem is that of
neutrino-less double beta decay, which is important in its own right.
There are also possibilities of doing geophysical and biological
experiments under extreme conditions, such as provided by an
under-ground laboratory.

\section{Some Ideas about the Detectors}

The following paragraphs summarise a few ideas on the concepts for a
new  detector guided by the physics discussions.

\subsection{Mega Water Cerenkov Detector (MWCD)}

One avenue for exploration we have discussed involves an extension of
the approach already demonstrated by the IMB, Kamioka and SuperKamioka
detectors, employing Cerenkov radiation from interactions or decays
in enormous volumes of water viewed by a surrounding surface of
photomultiplier tubes.  The largest of these, SuperK, has now been in
operation for 4 years, and is near saturation of its physics
capabilities---typically, further accumulation of data is expected to
only improve statistics of already-observed data. The SuperK detector has
approximately 10 thousand photomultipliers and a contained volume of 33
kilotons, and fiducial volume (2 m all around) of 22 kT.  The threshold
is several MeV.  We can gain greatly by raising the threshold of a
proposed detector to about 10 MeV, and reducing the percentage area
coverage from 40\% for SK, to 2\%, similar to that employed in IMB.  In
fact the IMB detector had a threshold of approximately 10 MeV. This
then leads to a gain of approximately a factor of 20 in the dominant
cost of photomultipliers and electronics.

The second area for improvement we have considered is in terms of
size.  By making the linear dimensions larger by a factor of two from
SuperK, we gain more than as the cube in fiducial volume: we would get
a 220 kT volume for nucleon decay, and 270 kT for SN neutrino
detection.  Five of these units would then yield 1.1 MT for nucleon
decay and 1.35 MT for SN.

This detector might be put in a long tunnel of about 80 m diameter and
400 m length. Orienting this array to point towards a neutrino factory
might be beneficial as well (or towards a local possible source of
neutrinos such as the galactic center).  One detector might be
constructed, with the others following in a few years.  An existing
railway tunnel might, for example, be used to provide initial access.
We estimate that the cost of the optical detectors, electronics and
infrastructure (water filtration) would be less than \$10M per module.
Excavation costs and those of tank (or cavity liner) and mechanical
structure, remain to be studied, as these depend heavily upon
location, type of rock and access.

This detector would answer every one of the high profile physics issues
we have discussed---all of which require data from an instrument in the
megaton class. This would, of course, not address the issue of solar
neutrinos below 10 MeV, and terrestrial neutrino detection; however, we
feel that that the solar neutrino area below 10 MeV will be well
explored by the time this instrument would come into operation, at
least ten years from now.  This detector would be the world's best in
nucleon decay, supernova neutrino detection, atmospheric neutrino
measurement, long baseline neutrino factory detection of muon
neutrinos, searches for point sources of neutrinos in the energy range
below about 1 TeV, searches for neutrinos from GRBs, and measurements
of the total neutrino flux from the sum of all supernovae throughout
the universe.  Note that a megaton instrument can easily detect the
initial burst of neutrinos from the in-fall stage of stellar collapse,
out through the whole galaxy.

The group also identified the following topics for further study with 
respect to the Mega Water Cerenkov Detector:
(1) Search for suitable location in consultation with mining engineers.
It could be an existing tunnel or a mine. The detector needs a cavern
of approximate length 400 meters with a cross section of 30 meters across.
(2) Study also the excavation and cavity lining techniques.
(3) Look into stimulation of PMT manufacture in India with possible 
cooperation from international companies. Also seek alternatives to PMT's.
(4) Look into water purification.
(5) Set up software Monte Carlo and study the performanance.
(6) Pursue studies of all physics goals with respect to the MWCD.
(7) The need to send some physicists to work at other neutrino labs in
order to gain experience.
(8) Seek international cooperation and collaboration for the project 
with an Indian lead.
 
\subsection{An Underground Neutron Detector for SN Neutrinos}

A possible proposal that can be built with the available expertise,
and also possibly available material, is a supernova detector. The
detector is in this case an array of neutron counters, possibly spread
over a large volume, by drilling holes in either a mountain base or a
mine with sufficient shielding. This process employs an enormous
amount of earth (megatons) as a neutron source.  The neutrons are
produced when neutrinos with energies above 20 MeV interact with the
surrounding rock. Both supernova explosions (short pulse) and
integrated neutrino events from a large number of supernovae may be
tracked with such a detector.

For a supernova explosion, this may infact be a real time detector. With
an appropriate design, such a detector can be combined to measure
penetrating charged particle fluxes (e.g. through-going upward muons from
astrophysical neutrinos) and for proton decay experiments (Kaon decay mode
which is complementary to that measured in water Cerenkov detectors) with
a denser array of the detector in part of the experiment.  Most worrisome
for this experiment are the natural neutron fluxes in the earth due to
radioactive decays.  The limiting sensitivity of such an experiment
depends upon locating very low radioactivity rock in a sufficiently deep
location.  This requires further investigation in India, and costs need
study but may be substantially less than a Cerenkov detector.

\subsection{DAEDULUS: Space Based Neutrino Detector}

This project is somewhat far into the future since it would require an
active space program if it is to be adopted for INO, but it is very
interesting.  Basically it is an idea for a neutrino detector
in space. Consider placing a neutrino detector with enough
shielding (from heat, background, etc) in a highly elliptic orbit
around the Sun. One could then measure the neutrino flux as a function
of the distance from Sun. Decent statistics would require the space
based neutrino detector to be in orbit for a sufficient amount of
time. It provides a unique method to understand neutrino oscillations,
which may be crucial in the instance of several scenarios of
oscillations not now resolved.

One can also watch for supernovae neutrinos and gamma-ray bursts, getting
accurate location by timing relative to terrestrial detectors (as is now
done for gamma ray detectors).  Further potential applications include
measurement of the ratio of pp to pep neutrino fluxes from the Sun, study
of cosmic ray isotropy and spectrum over one AU, and the study of possible
g-mode effects in cosmic rays. 

Background neutrino effects are taken care of since there is no
background from  terrestrial radioactivity, no reactors in the
vicinity, and no atmospheric cosmic ray muons and neutrinos. However
charged and neutral cosmic rays form a real background.

The techniques involved are similar to the LENS proposal. The difference
is that here it is proposed to isolate pure $^{176}$Yt (about two tons
of this gives the same sensitivity as LENS with 200 tons)). Such a
compact detector can be flown in some thing like a shuttle. Scaling
rates up from the present estimates for LENS there will be about, $g
\times 300$ (pp events), $g \times 160$ (Be events), $g \times 15$ (pep
events) per year.  Here $g$ is the gain from the distance factor
depending on how close the detector can get to the Sun. We aim for a
$g$-factor of 100, but even if $g \sim $ five to ten, detecting pep
neutrinos becomes a real and exciting possibility.

\section{Manpower requirements}

\begin{itemize}

\item
A chief problem area is the availability of a sufficient number
of trained physics personnel in India.  With declining number of
students getting into the science stream, and with the numbers going
down in the intake of Ph.D's in various institutes and universities, a
large experimental facility is bound to have manpower problems.  The
initiative must have enough breadth to be able to produce sufficient
number of Ph.D's, even though this is not a guarantee that students
will actually join such a project.  The very presence of such a world
class project would surely attract many talented young people.

\item An effort like INO would require large numbers of people
involving both research Institutes and Universities and probably funded
by many funding agencies.

\item Furthermore, the very nature and magnitude of the proposal also
demands international collaboration in order to be fully successful.

\end{itemize}

\subsection{Summary}

We have listed various points that were raised during four sittings
and many less formal discussions. While difficulties are many, there
are also attractive physics possibilities. A few stand out:

\begin{enumerate}

\item Initiation of studies and design, and ultimately construction of
a large new neutrino facility starting now blends well in time with
the exploitation of present instruments and is in synchrony with the
construction of a possible (probable, according to many) neutrino
factory.

\item  A strong motivation for locating a detector in India is the
near equatorial location, permitting observation of neutrinos passing
through the earth's core. The detector may be useful for  precision
studies like fixing the full MNS matrix (the equivalent of the CKM
matrix for neutrinos), etc.  In any case the location of a detector
near the equator provides a view that sweeps the entire celestial
sphere once per day for upcoming neutrinos.

\item The giant water Cerenkov detector goes a long way in addressing
many of the physics goals mentioned above.  At the megaton scale such
an unprecedentedly sensitive instrument will yield  good statistics at
reasonable cost while only trading off low threshold solar neutrino
physics (probably well studied by the time this experiment operates
anyway). The greatest advantages are, (1) that it will give a  good
understanding of the higher end of the solar neutrino spectrum, (2) it
can survey supernova neutrinos at megaparsec scale as well as measure
SN events inside our galaxy with great precision, (3) it will serve as
an excellent proton detector, and (4) it can also be used in a  long
base line experiment. In summary it may address all the physics issues
outlined here and probably many more.

\end{enumerate}

In short, we have identified some important and worthwhile physics
opportunities. Each of these needs to be examined more carefully with
the INO in view. While it's premature to expect that any  funding
proposal could come out from the short time-duration of the WHEPP-6
meeting, it is hoped that the group has identified viable concepts
for further investigation over the next few months. At that stage, it
may be pertinent to undertake a more detailed feasibility study for
such a detector. 

\section*{Appendix A: Solar neutrino event rates}

The solar neutrino rates for a 1 kTon water-Cerenkov detector are shown
in Table 1.  It is seen that the event rate falls sharply with
increasing threshold.

\begin{table}[hb]
\begin{tabular}{cc}
\hline
   $E_{min}$ (MeV) & Event rate \\ \hline
   8.0 &  .90   \\
  10.0 &  .25  \\
  12.0 &  .03 \\ \hline
\end{tabular}
\noindent\caption{Event rates per day at a 1 kTon water Cerenkov detector from solar neutrinos.}
\end{table}

\section*{Appendix B: Supernova events}

The number of supernova events during stellar collapse  are determined by 
the interaction of all
flavours (and antiflavours) with the material of the detector. For a
water-Cerenkov detector, this includes scattering off electrons, protons
and oxygen in the water. These give events that are forward-peaked,
isotropic and backward peaked respectively. Table 2 shows the events
accumulated by a kiloton detector from events from a supernova 10 KPc
away. There are few oxygen events without neutrino oscillations.
However, assuming a 3-flavour oscillation scenario consistent with solar
and atmospheric neutrino data, we see a significant enhancement (by a
factor of 6) of the oxygen events. This is because these events are
enhanced by mixing of the heavier flavours into the electron neutrino
spectrum. The hotter spectrum then gives a large event rate which is
strongly backward peaked. Increasing the threshold of
the detector from 8 to 12 MeV severely decreases the forward rate (by 
30 and 45\% with and without oscillations, while there is only a
marginal decrease in the isotropic rates (10\%) or the backward
rates (few \%) in the presence of oscillations.
\begin{table}[hb]
\begin{tabular}[b]{cccccccccc} \hline
$E_{min}$ (MeV) & \multicolumn{6}{c}{Events per kTon} &
\multicolumn{3}{c}{Ratio, $R = N/N_0$} \\
 & $N_0^F$ & $N_0^I$ & $N_0^B$ & $N^F$ & $N^I$ & $N^B$ & $R_F$ &
 $R_I$ & $R_B$ \\ \hline
    8.0  & 4.5 & 281.2 &   5.0  &  8.0 & 329.1 &  32.7 & 1.8 & 1.2 & 6.5 \\
   10.0  & 3.3 & 270.1 &   4.9  &  6.7 & 320.8 &  32.5 & 2.0 & 1.2 & 6.7 \\
   12.0  & 2.4 & 254.8 &   4.7  &  5.6 & 308.9 &  32.2 & 2.3 & 1.2 & 6.9 \\
   \hline 
\end{tabular}
\caption{Number of forward, backward and isotropic ($i = F, B, I$) events
with ($N^i$) and without ($N_0^i$) oscillations for a
supernova explosion at a distance of 10 KPc, for a 1 kTon detector. The
oscillation is assumed to be maximal.}
\end{table}

\end{document}